\documentclass[fontsize=12pt,pdftex]{scrartcl}

\usepackage[T1]{fontenc}
\usepackage{lmodern}
\usepackage{amssymb,amsmath}
\usepackage{mathtools}
\usepackage{autobreak}
\usepackage{comment}

\usepackage{braket}

\usepackage{authblk}

\title{Open string fluctuations around $KBc$ mulibrane backgrounds}
\author{Syoji Zeze}
\affil{Yokote High School, 68, Mutsunari, Yokote, Akita, Japan}
\date{23 September, 2024}

\begin{document}

\maketitle

\begin{abstract}
We investigate open string fluctuations around multibrane solutions within the $KBc$ subspace of open string field theory. Specifically, we consider space-time independent fluctuations of leading order zero in the $1/G$ expansion.   A constraint
on fluctuations around a multibrane solution is explicitly derived from the equation of motion. 
\end{abstract}

\section{Introduction}

The $KBc$ subspace~\cite{Okawa:2006vm} of open string fields has long been used to construct analytic solutions of the equation of motion (EOM) of open string field theory (OSFT).  The simple tachyon vacuum~\cite{Erler:2009uj} is a successful example of a ``$KBc$ solution'' in the early era.  Following this success, multibrane solutions were examined~\cite{Murata:2011ex,Murata:2011ep,Hata:2011ke,Hata:2012cy,Hata:2013hba}.   However, it turned out that the solutions do not reproduce the expected values of the gauge-invariant observables due to anomalies that arise from regularization of the singularity around $K=0$.  Quite later, anomaly-free solutions were constructed~\cite{Hata:2019dwu,Kojita:2019qbt}.  These solutions succeeded in avoiding anomalies by virtue of the extended degrees of freedom.

Despite these successful results, $KBc$ multibrane solutions have been known to exhibit an inconsistency.   That is, the solutions include divergent components in their Fock space expansion~\cite{Murata:2011ex,Murata:2011ep,Hata:2011ke}. This implies that an inner product between a solution and a certain fluctuation in the Fock space diverges and is therefore ill defined. Unfortunately, this inconsistency persists even in non-anomalus solutions~\cite{Hata:2019dwu,Kojita:2019qbt}. 

Deeper understanding the nature of open string flucuations around multiple D-brane backgrouds will help to resolve the inconsitency mentioned above.  Past studies have indicated that the space of open string field for a given reference boundary conformal field theory (BCFT) is unexpectedly vast enough to describe other BCFTs. A non-trivial background specifies fluctuations as a subspace of open string fields.  In fact, such a scenario is realized in the context of the construction of OSFT solutions in terms of the boundary condition changing operators~\cite{Erler:2014eqa,Erler:2019fye}.  In this context, orthogonal projectors define consistent subspaces of open string fluctuations for each boundary condition.  The multibrane solution constructed by this method naturally derives the Chan-Paton decomposition of open string fluctuations~\cite{Erler:2014eqa,Kishimoto:2014yea,Erler:2019fye}. 

We expect a similar scenario to the context of $KBc$ multibrane solutions; certain condition imposed on fluctations around a multibrane background will identify the consistent subspace of open string fields and removes the divergent components.  Unfortunately, a consistency condition that defines a space of fluctuations has not yet been identified in the $KBc$ context.  However, there is a nontrivial condition that should be imposed on fluctuations.  It is ``the EOM constraint'', the equation of motion contracted against a fluctuation.  The constraint is given by
\begin{equation}
   \mathcal{T}(\Phi, \Psi_N) = \int \Phi (Q_B \Psi_N +\Psi_N^2)  = 0, \label{eomtest}
\end{equation}
where $\Psi_N$ is a $KBc$ multibrane solution with brane number $N$, $\Phi$ is a fluctuation and $\int$ is the OSFT integral. The quantity $\mathcal{T}(\Phi,\Psi_N) $ has been called the EOM test~\cite{Hata:2019dwu}.  In past studies, the EOM test has been examined against a solution itself (that is, $\mathcal{T} (\Psi_0, \Psi_0)  $).    We note that the constraint \eqref{eomtest} should be imposed on any fluctuations other than the solution itself, as it ensures the nilpotency of the shifted kinetic operator $Q_B + \{\Psi_N, * \}$. 

At first sight, the constraint \eqref{eomtest} seems trivial, since $Q_B \Psi_N+\Psi_N^2$ is the left-hand side of the EOM. However, it is nontrivial since the solution breaks the EOM due to regularization of the singularity around $K=0$.  So-called $K_\epsilon$ regularization, namely replacement $K\rightarrow K+\epsilon$  yields
\begin{equation}
    \Psi_N^\epsilon = \psi_1^\epsilon  + \epsilon \psi_2^\epsilon, \label{soldecomp}
\end{equation}
where $\psi_1^\epsilon$ obeys EOM while  $\psi_{2}^{\epsilon}$ breaks it.  For the case $\Phi=\Psi_N$, it is known that the factor $\epsilon$ in the second term of \eqref{soldecomp} cancels the divergences due to $\Psi_1^\epsilon$  and $\Psi_2^\epsilon$ in the EOM test~\cite{Murata:2011ex,Murata:2011ep,Hata:2011ke,Hata:2012cy,Hata:2013hba,Hata:2019dwu,Kojita:2019qbt}.   However, the EOM test against arbitrary fluctuation $\Phi$ has not yet been examined systematically.~\footnote{Ref.~\cite{Hata:2019dwu} reported that the unitary transformed Fock vacuum passes the EOM test.   Ref.~\cite{Miwa:2020jps} studied the EOM tests between solutions with different brane numbers.}

In this study, we evaluated the EOM tests between a multibrane solution $\Psi_N$ and a scalar field $\Phi_M$ in the $KBc$ subspace.  We employed Hata's solution~\cite{Hata:2019dwu} with the ``brane number'' $N$\footnote{The solution represents $N+1$ branes since $N$ counts a number of branes added to the ``reference'' brane.} as a multibrane background.  We investigated fluctuations $\Phi_M$ with the brane number $M$ of the form 
\begin{equation}
  \Phi_M = H_{abc} (cK)_{ab} (Bc)_{bc}  
\end{equation}
\begin{equation}
H_{abc} = (G_a G_c)^\frac{M}{2}
\sum_{k=0}^M \sum_{l=0}^M
\frac{\beta_{kl}}{G_a^k G_b^{M-k-l}G_c^l} + \mathcal{O} \left(\frac{1}{G^M} \right) \label{osf}
\end{equation}
where we employed the notation of~\cite{Hata:2019dwu}. $H_{abc}$ depends on $K$ through the singular coordinate $G(K) = (1+K)/K$.  The complex coefficients $\beta_{kl}$ comprise a Hermitian matrix of dimension $(M+1)\times (M+1)$. Applying $K_\epsilon$ regularization to $\Phi$, we evaluated the EOM test in a way similar to the case $\Phi= \Psi_N$.   Then the EOM test gives a sum of the components of the matrix $\beta$.  Setting this to zero identifies the consistent subspace of open string fields around the background $\Psi_N$.  The main result of this study is the explicit presentation of the constraint that defines a consistent subspace around a multibrane background.  A constraint for any $(M,N)$ can be obtained with the help of a computer algebra system such as \textit{Mathematica}, as long as the computational resource allows.  While the dimension of a fluctuation increases as $(M+1)^2$, the EOM test reduces its dimensions only by one.   Thus, fluctuations are absent enough to develop OSFT around a multibrane background. 

This article is organized as follows. Section 2 introduces the basic ingredient necessary for the evaluation of the EOM tests.  Next, we explicitly evaluate the EOM tests for various values of $(M,N)$ in Section 3.  Finally, we discuss implications of our result in Section 4.
 
\section{EOM test}
\subsection{String fields and Equation of motion} 
Throughout this paper, we deal with classical string fields within the $KBc$ subspace. This means that any field has ghost number one. A string field in this setting is conveniently expressed in the ``tensor like'' notation  introduced in~\cite{Hata:2019dwu},  
\begin{align}
    \Phi & = H_{abc} (cK)_{ab} (Bc)_{bc} \\
        & =  H_{123} (cK)_{12} (Bc)_{23},  \label{ghost1}
\end{align}
where each subscript stands for position of operator insertion on the open string worldsheet in sliver frame.  $H_{abc}=H(K_a, K_b, K_c)$ is a three-dimensional function of $K$.\footnote{This can be understood as operator insertions in terms of suitable expansion.  For example, $H_{123} = \sum_{i} \alpha^{i}_{lmn} (K_1)^l (K_2)^m  (K_3)^n$ amounts to $\Psi = \sum_{i} \alpha^{i}_{lmn} K^l (cK) K^m (Bc) K^n $ in the standard notation of~\cite{Okawa:2006vm}} We will deal with real string fields. The reality condition for a real field reads $H_{abc} = H^{*}_{cba}$.  

Next, we consider a solution of the EOM $Q_B \Psi + \Psi^2 = 0 $.  A solution is expressed by
\begin{equation}
    \Psi=E_{abc}(cK)_{ab} (Bc)_{bc}
\end{equation}
according to~\eqref{ghost1}.  Plugging this into the left-hand side of the EOM, we obtain a ghost number two quantity $X_{abcd} (cK)_{ab} (cK)_{bc} (Bc)_{cd}$, where 
\begin{equation}
    X_{abcd} =  E_{acd}-E_{abd}
    + E_{abb} E_{bcd}-E_{abc}
    E_{ccd}
\end{equation}
Thus the EOM should be
\begin{equation}
    E_{acd}-E_{abd}
    + E_{abb} E_{bcd}-E_{abc}
    E_{ccd} =0, \label{EOM}
\end{equation}
up to an ambiguity which is irrelevant for our discussion.\footnote{
$X_{abcd}$ cannot be simply set to zero because of an ambiguity in the second ``b'' slot. Since $c^2=0$, the difference in the second slot by (const.)/$K$ leaves nothing.  Thus, the component equation should be $X_{abcd}=\frac{Y_{abc}}{K_b}$ where $Y_{abc}$ can be arbitrarily chosen. } Any solution of ~\eqref{EOM} can be parameterized by ``lower-rank'' objects $F_{ab}$ and $\Gamma_a$, which are obtained by assuming a pure gauge form of the solution or equivalently, by contracting~\eqref{EOM}. Either of them results in the following expressions.
\begin{equation}
    E_{abc} = F_{ac} + F_{ab} \frac{1}{\Gamma_b^2} F_{bc},
\end{equation}
where
\begin{align}
    \Gamma_{a}^2 & = \frac{1}{E_{aaa}} -1,\\ \label{Gamma_E}
    F_{ab} & = E_{abb} \Gamma_{b}^2\\
       & = E_{abb} \left( \frac{1}{E_{bbb}} -1   \right). \label{F_E}
\end{align}
The reality condition of $E_{abc}$ reduces to $F_{ab}=F_{ba}^*$ and $\Gamma^*_a=\Gamma_a$. 

\subsection{Multibrane solutions}
Here, let us describe the construction of Hata's multibrane solutions~\cite{Hata:2019dwu}.  One of the important feature of the solution is that component $E_{abc}$ is described by the ``singular coordinate''
\begin{equation}
    G(K) = \frac{1+K}{K}.
\end{equation}
The singularity of this coordinate at $K=0$ plays an important role in defining a non-trivial solution.  It is also important to note that $1/G$ defines a regular coordinate that will be used in the $K_\epsilon$ regularization scheme.  

Another important feature of the solution is the brane number $N$. It counts the order of singularity through $\Gamma^2_a$.
\begin{equation}
    \Gamma_a^2 = G_a^N, \label{gamma}
\end{equation}
where $G_a =G(K_a)$.  Detailed analysis in \cite{Hata:2019dwu} has shown that the value of the OSFT action for the solution is given by $N$. This means that the solution represents the $N+1$ branes.  The candidate for $F_{ab}$ that satisfies $F_{aa}=1-\Gamma_a^2$, which follows form~\eqref{F_E} and~\eqref{Gamma_E}, has been proposed as
\begin{equation}
    F_{ab} = \prod_{k=1}^N 
    \left(1-G_a^k G_b^{N-k}\right)^{\alpha_k}.
\end{equation}
where $\alpha_k$ is a nonnegative constant that satisfies
\begin{equation}
    \sum_{k=1}^N \alpha_k = 1.
\end{equation}
The hermiticity condition $F^*_{ab} = F_{ba}$ is achieved by imposing $\alpha_{N-k}=\alpha_k$.   It has been shown that $\{\alpha_k\}$ is overcome by requiring anomaly cancelation on gauge-invariant observables~\cite{Hata:2019dwu,Hata:2019ybw}.  

\subsection{Evaluation of the EOM tests}
Let $\Psi_N$ be a multibrane solution with an index $N$, and consider a fluctuation $\Phi$ around the solution.  The EOM test for $\Psi_N$ against $\Phi$ is given by
\begin{equation}
    \mathcal{T}( \Phi, \Psi_N) 
    = \int \Phi (Q_B \Psi_N + \Psi_N^2).
\end{equation}
As explained in Introduction, this will not vanish identically due to $K_\epsilon$ regularization, which replaces \textit{all} $K$ with $K+\epsilon$ in string fields.   The regularized solution is given by
\begin{equation}
    \Psi_N = E^{\epsilon}_{abc} (c K_{\epsilon})_{ab}(Bc)_{bc}, \label{regsolution}
\end{equation}
where $E^\epsilon_{abc}=E(K_\epsilon,K_\epsilon,K_\epsilon)$ and $K_\epsilon=K+\epsilon$.  The solution can be decomposed into two terms, 
\begin{equation}
    \Psi_N = E^{\epsilon}_{abc} (c K)_{ab}(Bc)_{bc} +
    \epsilon E^{\epsilon}_{abc} (c)_{ab}(Bc)_{bc}.\label{decomp}
\end{equation}
For finite but small $\epsilon$, the second term of~\eqref{decomp} breaks the EOM, while the first term does not.  Then, using EOM for the first term, we obtain the following.
\begin{equation}
    Q_B \Psi_N +\Psi_N^2 =  \epsilon E^{\epsilon}_{abc} (c K_{\epsilon})_{ab}  c_{bc}.  
\end{equation}
Similarly to the EOM solution,  a fluctuation is decomposed into 
\begin{equation}
    \Phi =  H^{\epsilon}_{abc} (c K)_{ab}(Bc)_{bc} +
    \epsilon H^{\epsilon}_{abc} (c)_{ab}(Bc)_{bc}. \label{flucreg}
\end{equation}
Here, the component $H^\epsilon_{abc}$ does not need to obey the EOM.  Using \eqref{regsolution} and \eqref{flucreg}, we obtain an expression for the EOM test,
\begin{equation}
    \mathcal{T}(\Phi, \Psi_N) 
    = \epsilon  \int H^\epsilon_{123} E^\epsilon_{341}
    (cK_\epsilon)_{12}(Bc)_{23} (cK_\epsilon)_{34} c_{41}. \label{EOMtest}
\end{equation}
This can be effectively evaluated by expanding $H_{123} E_{341}$ in $1/G$ or $K$. 
\begin{align}
  H^\epsilon_{123} E^\epsilon_{341}& = 
  \sum_{\Sigma n_{k} = 0}  a_{n_1, n_2, n_3, n_4}  G_{1}^{n_1} G_{2}^{n_2} G_3^{n_3} G_{4}^{n_4}\\
  &= 
  \sum_{\Sigma n_{k} = 0}  a_{n_1, n_2, n_3, n_4}  K_1^{-n_1} K_{2}^{-n_2} K_3^{-n_3} K_{4}^{-n_4},
\end{align}
where $a_{n_1, n_2, n_3, n_4} $ is an numerical coefficient, and we used the fact that $G_\epsilon$ can be approximated to $1/K_\epsilon$ for small $\epsilon$.  This expansion enables us to express the EOM test by the fundamental correlator $\mathcal{T}_{n_1,n_2, n_3,n_4}$~\cite{Hata:2019dwu}, 
\begin{align}
    \mathcal{T}(\Psi_N, \Phi) & =
    \sum_{\Sigma n_{k} =0}
    a_{n_1,n_2, n_3,n_4} \times \epsilon
    \int Bc \frac{1}{K_\epsilon^{n_3}}
    c \frac{1}{K_\epsilon^{n_4-1}}
    c \frac{1}{K_\epsilon^{n_1}}
    c \frac{1}{K_\epsilon^{n_2-1}} \\
    & = \sum_{\Sigma n_{k}=0} 
    a_{n_1,n_2, n_3,n_4} \mathcal{T}_{n_1,n_2, n_3,n_4}.\label{EOMtestfinal}
\end{align}
The fundamental correlator was evaluated in~\cite{Hata:2019dwu} with the help of various techniques.  It can be obtained from
the $Bcccc$ correlator on the cylinder of the sliver frame, which includes multiple integrals over the width of strips inserted between $c$s.   The $s-z$ trick~\cite{Murata:2011ex,Murata:2011ep,Hata:2011ke} is used to evaluate multiple integrals. The explicit value of $\mathcal{T}_{n_1,n_2, n_3,n_4}$ has already been given in terms of the confluent hypergeometric function~\cite{Hata:2019dwu}. 

\subsection{String fields in the $K_\epsilon$ regularization scheme}
In the $K_\epsilon$ regularization scheme, the singular coordinate $G(K)$ is regarded as a $1/\epsilon$ quantity according to
\begin{equation}
    G(K_\epsilon) = \frac{K+1+ \epsilon}{K+\epsilon} \sim \frac{1}{\epsilon}
\end{equation}
for small $K$.  Thus, the components $E^{\epsilon}_{abc}$ or $H^{\epsilon}_{abc}$ will be expanded in $1/G\sim \epsilon$.  First, the $1/G$ expansion of $E{^\epsilon}_{abc}$ has already been derived from~\cite{Hata:2019dwu} to be 
\begin{equation}
    E^N_{abc}
    = (G_a G_c)^\frac{N}{2} \sum_{k=0}^N \alpha_k
    \left(
     \frac{1}{G_a^k G_c^{N-k}} -\frac{1}{G_a^k G_b^{N-k}} -
     \frac{1}{G_b^k G_c^{N-k}}
    \right)+ \mathcal{O} \left(\frac{1}{G^N} \right). \label{solexpanded}
\end{equation}
Here we introduce the notion of the ``order'' of a string field that counts the order of $G$ of each term in the expansion $1/G$. The order can be regarded as an eigenvalue of a derivation on string fields.   In terms of the Lie derivative introduces in~\cite{Hata:2021lqz}, a deviation $\mathcal{N}$ in the $KBc$ subalgebra can be defined.  The operation of $\mathcal{N}$ on each of the $KBc$ fields is given by
\begin{equation}
    \mathcal{N} K = K \frac{G}{G'},  \quad
    \mathcal{N} B = B \frac{G}{G'}, \quad
    \mathcal{N} c = - c\frac{G}{G'} B c.
 \end{equation}
For a ghost number 1 string field of the form 
\begin{equation}
    \Phi= G_a^m G_b^n G_c^l  (cK)_{ab}(Bc)_{bc}
\end{equation}
we have $\mathcal{N}\Phi = (n+m+l) \Phi$. Thus, the order is assigned to each term of the expansion~\eqref{ghost1}. Note that the leading order of \eqref{solexpanded} is 0.  This feature is crucial for the finiteness of the observables.

Having described the background solution, we shall present a candidate of a fluctuation.  First, we consider fluctuations with leading order zero.   Next, we employ an expansion similar to~\eqref{solexpanded} in which the sum of order $G^{-N}$ terms in the round brace can be chosen arbitrary.  We also require the fluctuation to be real. A candidate for such fluctuation with a factor $(G_a G_c)^\frac{M}{2}$ is 
\begin{equation}
    H^M_{abc} = (G_a G_c)^\frac{M}{2}
    \sum_{k,l=0}^{M} \frac{\beta_{k,l}}{G_a^{k} G_{b}^{M-k-l} G_{c}^{l}} + \mathcal{O} \left(\frac{1}{G^M} \right),\label{HM}
\end{equation}
where $\beta_{kl}$ is a component of a $(M+1)\times(M+1)$ Hermeite matrix whose real dimension is $(M+1)^2$.  The EOM test for a solution $\Psi_N=E^N_{abc} (cK)_{ab}(Bc)_{bc}$ and a fluctuation $\Phi_M = H^M_{abc} (cK)_{ab}(Bc)_{bc}$ can be evaluated according to~\eqref{EOMtestfinal}.  As $\mathcal{T}_{n_1,n_2, n_3,n_4}$ is already given, we only need to extract $a_{n_1,n_2, n_3,n_4}$ from a product of $E^N_{abc}$ and $H^M_{abc}$. We shall call $N$ and $M$ the ``brane numbers'' of the background and a fluctuation, respectively.  The brane number is characterized by the exponent of the $(G_aG_c)^{M/2}$ factor in each string fields.

\section{EOM constraint on fluctuations}
Here we evaluate EOM tests for the explicit choices of $N$ and $M$.  In principle, $N$ and $M$ can be any positive integers. However, we restrict ourselves to the cases $M+N$ even.  The reason for the restriction is that the $(G_a G_c)^{\frac{N+M}{2}}$ factor in the integrand of the EOM test introduces brunch cuts on the $z$ plane and requires a choice of integration contour.  We simply avoid this by setting $M+N$ even, for convenience.   According to~\eqref{EOMtestfinal}, a value of the EOM test is given by linear combination of $\beta_{k,l}$ 
\begin{equation}
    \mathcal{T}(M,N) = \sum_{k,l=0}^{N} \omega_{k,l} \beta_{k,l}
\end{equation}
where $\omega_{k,l}$ is numerical coefficients.  Hereafter we abbreviate $\mathcal{T}(\Phi_M, \Psi_N)$ to $\mathcal{T}(N,M)$ .

First, we set $M$ equal to $N$. This can be regarded as a ``modest'' fluctuation since it includes the background solution as a special case.  $N=0$ is trivial since this corresponds to the perturbative vacuum.  The next example of $N=1$ corresponds to double brane.   In this case, it turns out that the EOM test vanishes trivially.
\begin{equation}
\mathcal{T}(1,1) = 0. 
\end{equation}
The trivial EOM test for $N=1$ has already been observed in the analysis of~\cite{Hata:2019dwu}. This means that the coefficients for the fluctuation $\beta_{00}, \beta_{01}, \beta_{11}$ can be arbitrarily chosen.   For $N=2$, we obtain a nontrivial result
\begin{equation}
    \mathcal{T}(2,2) = -2 \beta_{00}.
\end{equation}
Setting the requirement of the vanishing EOM test set $\beta_{00}$ to zero.   Thus, the number of remaining components of $\Phi_2$ is $3^2-1=8$.   As $N$ increases,  the constraint equation involves more components.  For $N=3$ and $4$, we obtain
\begin{equation}
    \mathcal{T}(3,3) = -\frac{2}{3} \left(28 \beta _{00}+29 \beta _{01}+11 \beta _{02}-11 \beta _{03}+10 
   \beta _{11}+11 \beta _{12}-11 \beta _{13}+\beta _{22}-3 \beta _{23}\right),
\end{equation}
\begin{equation}
    \mathcal{T}(4,4) = \left(\frac{4 \pi ^2}{3}-10\right) \beta _{00}-2 \left(6 \beta _{01}-\beta
   _{02}+\beta _{03}-\beta _{04}+2 \beta _{11}+\beta _{13}-\beta _{14}+\beta
   _{23}+2 \beta _{24}\right).
\end{equation}
These constraints remove a single component from $\Phi_M$ .  Constraint eqautions for larger $N$ can be obtained similarly.  

Next, we consider the cases where $M$ and $N$ are different.  As an example, we will fix $N$ to 1 and increase $M$. As we have already seen, the first nontrivial case, $\mathcal{T}(1,1)$ is zero.  The next is 
\begin{equation}
    \mathcal{T}(3,1) = 4 \left(2 \beta _{00}+\beta _{01}-\beta _{03}-\beta _{13}\right)
\end{equation}
Increasing $M$ further, we obtain
\begin{align}
\begin{autobreak}
   \mathcal{T}(5,1) 
   = -8 \bigl( 2 \pi ^2-3\bigr) \beta _{00}
   -\frac{4}{3} \Bigl(\bigl(4 \pi ^2-21\bigr)
   \beta _{01}
   -6 \beta _{02}
   +3 \beta _{04}
   +12 \beta _{05}
   -6 \beta _{11}
   -3 \beta_{12}
   +3 \beta _{14}
   +12 \beta _{15}
   +3 \beta _{24}
   +6 \beta _{25}
   -2 \pi ^2 \beta_{55}\Bigr)
   \end{autobreak}
\end{align}
\begin{align}
    \begin{autobreak}
       \mathcal{T}(7,1)=
       \frac{4}{15} \bigl(4 \bigl(45-130 \pi ^2+4 \pi ^4\bigr) \beta _{00}
       +\bigl(255-390\pi ^2+4 \pi ^4\bigr) \beta _{01}
       -50 \bigl(2 \pi ^2-3\bigr) \beta _{02}
       -5\bigl(2 \pi ^2-9\bigr) \beta _{03}
       -5 \bigl(3+2 \pi ^2\bigr) \beta _{05}
       -20\bigl(3+\pi ^2\bigr) \beta _{06}
       +5 \bigl(2 \pi ^2-27\bigr) \beta _{07}
       -30\bigl(2 \pi ^2-3\bigr) \beta _{11}
       -5 \bigl(4 \pi ^2-21\bigr) \beta _{12}
       +30\beta _{13}-15 \beta _{15}
       -60 \beta _{16}
       +15 \bigl(2 \pi ^2-9\bigr) \beta_{17}
       +30 \beta _{22}
       +15 \beta _{23}
       -15 \beta _{25}
       -60 \beta _{26}
       +5 \bigl(4\pi ^2-21\bigr) \beta _{27}
       -15 \beta _{35}
       -30 \beta _{36}
       +5 \bigl(2 \pi^2-9\bigr) \beta _{37}
       +10 \pi ^2 \beta _{57}
       +10 \pi ^2 \beta _{66}
       +100 \pi ^2\beta _{67}
       -2 \pi ^2 \bigl(\pi ^2-90\bigr) \beta _{77}\bigr)
    \end{autobreak}
\end{align}
A similar calculation can be performed for any pair of positive integers $M$ and $N$ as long as $M+N$ is even, and likely to give a non-trivial EOM constraint.   

In closing this section, we make a comment on a direct sum of open string fluctuations with different brane numbers.  Note that fluctuations with different brane numbers are not distinguished by the eigenvalue of $\mathcal{N}$ --- they all are order zero.   Therefore, we can consider a fluctuation such as $\Phi_1 + \Phi_2 + \Phi_3 + \cdots$, where $\Phi_k$ denotes a flucutuation with brane number $k$.   The EOM test for such a sum gives an EOM constraint.

\section{Discussions}
In this study, we investigated the EOM constraints on open string fluctuations around multiple $N+1$ branes.  We considered open string fluctiations with leading order zero in the $1/G$ expansion. The brane number $M$ of a fluctuation was introduced.  The EOM constraints were obtained for even values of $N+M$.   We find that the EOM constraint eliminates a single component among the $(M+1)^2$ fluctuations for each brane number.   This defines $(M+1)^2-1$ consistent fluctuations for each number of brane.     

Our motivation for this study is to identify a consistent subspace of open string fluctuations among the full space of open string fields.  As we have already shown, the EOM test against fluctuations with no momentum is nonzero in general. On the other hand, the EOM test against fluctuations with nonzero momentum always vanishes. This is due to momentum conservation.  Since the left hand side of EOM carries no momentum,  the EOM test against a fluctuation vanishes identically,
\begin{equation}
    \int (Q_B \Psi_N + \Psi_N^2) \Phi(p) = 0,
\end{equation}
where $\Phi(p)$ is a fluctuation that carries momentum.  Therefore, the EOM does not impose any constraint on momentum fluctuations. However, we can still require $\Phi(p)$ to have a smooth limit to a fluctuation $\Phi$ that has vanishing EOM test: 
\begin{equation}
    \lim_{p\rightarrow 0} \Phi(p) = \Phi.
\end{equation}
Thus, the above condition gives a nontrivial constraint on fluctuations with nonzero momentum.  A candidate for such a fluctuation will be obtained by inserting an off-shell vertex operator into a fluctuation with no momentum.  Such a construction will be an important step towards complete specification of open string fluctuations.  

Another aspect of open string fluctuations we would like to discuss here is the relation between them.  One motivation for this study is the Chan-Paton decomposition. In order to identify orthogonal fluctuations, we have to evaluate OSFT integrals such as $\int \Phi_1 Q_{\Psi_N} \Phi_2$ or $\int \Phi_1 \Phi_2 \Phi_3$.   Both of their integrands takes form of
\begin{equation}
    U_{12345} (cK_\epsilon)_{12}(cK_\epsilon)_{34} (cK_\epsilon)_{45} (Bc)_{51}, 
\end{equation}
where $U_{12345}$ is a order zero object.  We find that this kind of product cannot be easily evaluated in terms of the method of~\cite{Hata:2019dwu}.  The product is expressed by an integral with respect to $s$ and $z$ due to the $s-z$ trick.  The $\epsilon$ dependence is completely absorbed by the same rescaling as the EOM case.  However, the $s$ integral gives a divergent factor 
\begin{equation}
    \int_{0}^\infty ds \frac{1}{s}e^{-s} = \Gamma(0). 
\end{equation}
This result is inconsistent with earlier results for the value of the classical action~\cite{Murata:2011ep,Hata:2011ke, Hata:2019dwu}, where such a kind of integrals were successfully evaluated.  A possibly solution to this problem would be the inclusion of subleading terms in the $1/G$ expansion.~\footnote{The evaluation of classical action for a EOM solution~\cite{Hata:2019dwu} avoids this difficulty in terms of ``topological'' argument.  The integral in question is reduced to the one with the same leading order as the EOM tests.  This procedure cannot be applied to our case, since flucutations are no more solutions of EOM. }
It will be important to develop an efficient method to evaluate OSFT integrals for arbitrary $KBc$ fields.  Such a development will be important to understand the nature of open string fields.

\section*{Acknowledgement}

The author thanks Yukihisa Sato for early stages of this work.  The author also thanks Hiroshi Kunitomo for a valuable discussion at the QFT2023 conference. 

\bibliographystyle{utphys}
\bibliography{hep}

\end{document}